\documentclass[aps,pre,preprint,showpacs,floatfix]{revtex4}
\begin{document}
\baselineskip 20 true pt
\preprint{APS}

\def\A{{\cal A}}
\def\F{{\cal F}}
\def\K{{\cal K}}
\def\IF{{\bf I_{\cal F}}}
\def\IA{{\bf I_{\cal A}}}

\title{Scaling of the Background Brain Dynamics and Alpha Ryhthm}

\author{D.C. Lin$^1$, A. Sharif$^1$, H.C. Kwan$^2$}

\affiliation{$^1$Department of Mechanical and Industrial
Engineering, Ryerson University, Toronto, Ontario, Canada\\
$^2$Department of Physiology, University of Toronto, Toronto,
Ontario, Canada}

\date{\today}

\bigskip

\begin{abstract}
The scaling property of the brain dynamics is studied based on the
zero-crossing of the local electroencephalographic (EEG) recording
taken from healthy young adults in eyes closed and eyes open.
Evidence of coupling between the EEG fractal dynamics and the
$\alpha$ rhythm is presented. An organization principle that
governs this coupling relationship is proposed. In the $\alpha$
dominant brain state, a possible interpretation using the
self-organized criticality similar to the punctuated equilibrium
is discussed.
\end{abstract}


\maketitle

\section{Introduction}
\bigskip

The cortical activity of the human brain in wakefulness and
eyes-closed typically exhibits the 8$\sim$12 Hz $\alpha$ rhythm$^1$.
Although its origin remain open$^{2,3}$, this classical rhythm has
been associated with the ``resting" state of the cortex with most
pronounced activity recorded in the parietal and occipital areas.

While the $\alpha$ rhythm represents one of the major cortical
activities, there typically lies a background ``noise" component
of unknown functionality and origin$^4$. There is growing evidence
of the importance of this background component. For example, local
surface scalp measurements based on electroencephalography$^{4,5,6,
7}$ (EEG) and magnetoencephalography$^8$ reveal its fractal
characteristics coexisting with moderate $\alpha$ rhythm. Similar
results were found in multi-channel recording$^{4,6,7,8}$,
suggesting the global nature of the fractal dynamics in the cortex.
The relevance of this background fluctuation in physiology may be
seen from its state dependence property$^{8,22}$ where normal
individuals in eyes closed average to a larger scaling exponent
than in eyes open. Recent works further imply that the EEG
fluctuation may have a broader implication on the other autonomic
function such as the cardiovascular regulation$^{9,10}$.

The purpose of this study is to investigate the EEG background
fluctuation and its coexistence with the $\alpha$ rhythm. While they
are separately subjects of intense study, far less understood is the
organization of these two prominent features of the brain dynamics.
In addition, the findings reviewed above may require further
clarifications when the $\alpha$ oscillation becomes dominant. For
example, a trained meditator or Yoga practitioner can shift a
significant portion of the EEG signal power to the $\alpha$ band that
obscures any potential fractal characteristics in the background
fluctuation. Fig.~1 shows EEG's with moderate and strong $\alpha$
rhythm from a normal subject and an experienced Yoga practitioner,
respectively. It is observed that the power law trend diminishes
almost entirely in the $\alpha$ dominant EEG, but not in the $\alpha$
moderate EEG. This apparently inverse relationship with the strength
of $\alpha$ rhythm implies either the fractal background fluctuation
is buried in the $\alpha$ oscillation, and can thus no longer be
detected from the amplitude characteristics of the EEG, or there is
simply no long-range correlated fluctuation in the $\alpha$ dominant
brain state. Using the detrend fluctuation analysis$^{11}$ (DFA) on
the integrated $\alpha$ dominant EEG actually supports the latter
scenario of a white noise process in the background fluctuation (DFA
exponent $\sim$0.5); see Fig.~1c. However, as we show below, it is
not a valid description for the $\alpha$ dominant brain state.

In this work, we use the zero-crossing property to study the EEG
background fluctuation coexisting with $\alpha$ rhythm. EEG
zero-crossing has been primarily used to extract event-related
frequency information$^{12}$. Despite the obvious advantage of
being insensitive to amplitude artefacts, the use of zero-rossing
to study EEG background fluctuation appears scarce. Watters and
Martins used DFA to analyze the ``EEG walk" constructed from the
zero-crossing of EEG showing moderate $\alpha$ rhythm$^{13}$. These
authors found evidence of scaling and rejected the (uncorrelated)
random-walk interpretation for the EEG background component. But
we should point out that a DFA scaling exponent $\sim 0.5$ (random
walk) is in fact possible in the $\alpha$ dominant brain state
(Fig.~1c).

By excluding the $\alpha$ wave zero-crossing, we will show that the
EEG background does exhibit fractal characteristics. For the scaling
analysis in $\alpha$ dominant case, the zero-crossing approach is
found to be more effective than other amplitude-based methods such
as the power spectral density function and DFA. The complementary
set of the background zero-crossing in the real line captures other
EEG activities. For the $\alpha$ dominant case, this complementary
set describes mainly the $\alpha$ dynamics. With the confirmed EEG
fractal background, we conjecture fractal scaling in the emergence
of $\alpha$ dynamics. Support to this conjecture is found in the
power law distribution of the $\alpha$ interval. Our analysis also
suggests that the $\alpha$ dominant brain state may be interpreted
in the unversality class of the self-organized criticality$^{15}$
of punctuated equilibrium$^{14}$. While SOC has been proposed for
the EEG fluctuation in the $\alpha$ frequency band$^{7,8}$, and its
analytical phase$^4$, our result provides further evidence of its
possible connection to the organization of the EEG background
dynamics and the $\alpha$ rhythm. Finally, we are able to write
down the governing principle for this organization that relates the
EEG fractal scaling and the emergence of $\alpha$ dynamics.

In the next section, the main idea of extracting the zero-crossing
of EEG background fluctuation is described and verified with
artificial examples. In section III, the method is applied to the
EEG records taken from subjects showing little to significant
$\alpha$ rhythm. Discussion and concluding remarks are given in
the last section.

\section{Methods and Numerical Examples}
\bigskip

\subsection{Main Ideas}

Let the EEG be $x(t)$. The zero-crossing time is the level set
$\{t_i,x(t_i)=0\}$ where the index $i$ registers the order of the
zero crossing event. In practice, $\{t_i\}$ is first determined
by linear interpolation and then used to define the set of
crossing-time-interval (CTI) ${\cal C}=\{\tau_i=t_{i+1}-t_i\}$.

Zero-crossing of a stochastic process is a surprisingly hard problem;
see, e.g., Ref.~16. Thanks to its self-similarity, the CTI for a
fractal process is known to follow a power law distribution$ ^{17}$:
$p(\tau)\sim\tau^{-\nu}$, where $p(\tau)$ is the probability density
function (PDF). For example, $\nu=2-h$ for the fractional Brownian
motion (fBm) $B_h(t)$ of Hurst exponent $h$.

If fractal exists in the $\alpha$ dominant EEG, it can be captured
in ${\cal C}\backslash{\cal A}_\alpha$ where ${\cal A}_\alpha$
denotes the CTI of the $\alpha$ oscillation. However, fractal CTI's
can occasionally lie in the $\alpha$ rhythm range$^{18}$. It is
thus not possible to obtain ${\cal A}_\alpha$ based solely on the
value of CTI.

What is characteristic to the rhythmic oscillation in general is a
steady zero-crossing pattern of the oscillation. Hence, a set of
CTI's is considered of $\alpha$ origin if they correspond to {\it
continuous} zero-crossing in the $\alpha$ rhythm range. Specifically,
we consider a bigger set $\A$ using the following two criteria: (i)
the CTI lies in the range of $\alpha$ rhythm range and (ii) the
CTI's are from continuous zero-crossing. Let such CTI's be denoted
by
$$A_i=\{\tau_{l_i},\tau_{l_i+1},\cdots,\tau_{l_i+m_i}\},
\eqno(1a)$$
$m_i>1, i=1,2,\cdots$. Thus, $A_i$ describes the $i^{\rm th}$
$\alpha$ event segment in the EEG record. The CTI's for the $\alpha$
dynamics is estimated by the union of $A_i$'s:
$$\A=\cup A_i.
\eqno(1b)$$
Inevitably, some continuous zero-crossing could still come from the
potential fractal component of the EEG. Hence, $\A_\alpha$ is only
a subset of $\A$. The ``error" $\A\backslash\A_\alpha$ depends on
the fractal property as well as the strength of the $\alpha$ rhythm.
The complement ${\cal C}\backslash\A$ captures irregular zero-crossing
that is characteristic to the fractal process

For moderate $\alpha$ oscillation, EEG may contain other rhythmic
components. The check of continuous zero-crossing must then be
extended to a reasonably large scale range. To this end, we first
consider the set of large CTI fluctuation: $I_1=\{\tau_i>\tau_u\
{\rm or}\ \tau_i<\tau_l\}$ where $\tau_u,\tau_l$ are the upper and
lower thresholds for defining the set of large CTI. Let $\K$
denote the CTI's from continuous zero-crossings in $({\cal C}
\backslash I_1)\backslash \A$ and let $I_0=({\cal C}\backslash I_1)
\backslash(\K\cup\A)$. Finally, the set of fractal CTI is obtained
by
$$\F=I_1\cup I_0
\eqno(1c)$$
Note that this definition does not preclude fractal crossing in the
$\alpha$ rhythm range. Note, for the $\alpha$ dominant case, that
continuous crossing is mainly captured in $\A$ and, thus, $\K\sim\{
\emptyset\}$ and we recover $\F\sim {\cal C}\backslash\A$ as shown
above. If $\tau_u\gg\mu(\tau)\gg\tau_l$, where $\mu(\tau)$ is the
mean of $\{\tau_i\}$, any accidental fractal crossing included in
$\K\cup\A$ will not introduce bias to the power law distribution of
the fractal crossing (Fig.~2 below).

\subsection{Numerical Examples}

To demonstrate the above idea, we used Hilbert transform to
construct the fractal time series given by the amplitude process
of fBm, $A_h(t)$. This is to mimic the fractal property reported
in the $\alpha$ band-passed EEG$^7$. Note that $A_h(t)$ inherits
the same scaling characteristics from $B_h(t)$. Hence, $h=2-\nu$
holds, where $\nu$ is estimated from the histogram of ${\cal C}
\backslash\A$. To define $\A$, $\tau_l\sim\exp(-5)$ and $\tau_u
\sim\exp(-1)$ were used. These values are determined from the
range of CTI of $A_h(t)$: $\tau_u\sim 0.8\max({\cal C})$, $\tau_
l\sim\Delta t$, the sampling time. Fig.~2 shows the PDF estimate
of ${\cal C}\backslash\A$. It is seen that theoretical $\nu$
values are verified {\it before} and after deleting $\A$ (Fig.~2).
This should be the case since no band-limited component exists
in $A_h(t)$. It thus establishes deleting $\A$ defined above does
not affect the power law PDF of a pure fractal signal.

To examine the influence from the band-limited rhythmic oscillation,
$A_h(t)$ in randomly selected time intervals of variable length
were replaced by a narrow-band process $x_\alpha(t)={\cal M}(t){\cal
N}(t)$ where ${\cal M}=1+A_h(t)$ models the fractal amplitude
modulation$^7$ and the narrow-band ${\cal N}(t)$ is a sine wave of
Gaussian amplitude ${\bf X}$ and frequency ${\bf f}$; i.e., ${\bf X}
={\bf N}(1,\sigma_{\bf X})$ and ${\bf f}={\bf N}(10,\sigma_{\bf f})$.
Note that ${\cal N}(t)$ has a 10-Hz central frequency to mimic the
$\alpha$ rhythm. To simulate the dominance of the band-limited
oscillation, the probability of an interval being selected for $x_
\alpha(t)$ is four times of those for $A_h(t)$. In addition, the
interval length for $x_\alpha(t)$ is at least three times shorter
than those for $A_h(t)$. The synthetic data so constructed is shown
in Fig.~3a.

In presenting our results, we keep the time unit in all figures so
as to make easy reference to the narrow-band oscillation. Segments
of CTI's before and after deleting $\A$ are shown in Figs.~3b, 3c.
The corresponding $p(\tau)$ are estimated in Fig.~3d. It is clear
that the narrow-band component $x_\alpha$ can create significant
bias in the otherwise power law PDF of $A_h(t)$. The theoretical
power law $p(\tau)$ for $A_h(t)$ is correctly described after
deleting $\A$ defined by (1).

\section{EEG Scaling and the Alpha Dynamics}
\bigskip

\subsection{Scaling of the EEG Background Fluctuation}

We now apply the zero-crossing method to EEG records with varying
degrees of $\alpha$ rhythm. These records were collected from six
subjects (3 males, 3 females) of age 21 to 30 year-old (mean:
$\sim$24 who gave written consent to participate in the study. All
subjects were instructed to maintain normal daily activity before
participating in the 5-minute recording session. Surface scalp
electrodes were attached according to the 10-20 international
system at O1, O2, referencing to Cz. Two groups of data were taken:
one in eyes open (EO) and one in eyes closed (EC). For EO,
subjects were asked to direct their gaze at certain part of a
shielded room to minimize eyes movements. For EC, no specific
instruction was given to the subjects other than to relax and have
their eyes closed. Output impedence from the recording system has
been kept below 5k$\Omega$. The EEG was first band-passed from 0.1
to 70 Hz and then digitized with a 12-bit A/D precision at 250 Hz
(first four subjects) and 500 Hz (last two subjects).

In order to measure the strength of the $\alpha$ rhythm, we use the
ratio of EEG signal power in the 8$\sim$12 Hz band to the full
accessible frequency range, $R_\alpha=\int_8^{12}S(f) df/\int S(f)
df$ (Fig.~4a). As expected, the $R_\alpha$ is larger in EC than in
EO due to the lack of visual stimulation in EC. Three of the six
subjects (S2, S3, S4) are able to generate dominant $\alpha$ rhythm
with large $R_\alpha$ measure ($>0.45$) in EC. It is important to
note that low $R_\alpha$ measure could mean very little $\alpha$
activity; e.g., the corresponding power spectra are given by power
law with little identifiable feature in the $\alpha$ band. In
contrast, there is always a very distinct ``$\alpha$ peak" located
at the 10Hz range for subjects showing large $R_\alpha$.

The CTI PDF's of all EEG data sets are found to be of the power law
form $p(\tau)\sim\tau^{-\nu}$. This indicates the fractal dynamics
continues to exist in the brain state showing $\alpha$ rhythm. In
particular, qualitatively different $p(\tau)$'s are found before and
after deleting ${\cal A}$ from subjects showing dominant $\alpha$
rhythm (Fig.~4b), and almost the same $p(\tau)$ from subjects showing
moderate to little $\alpha$ rhythm (Fig.~4c, also Fig.~2). For
estimating the scaling in $\alpha$ dominant brain state, this means
a more effective approach of using EEG zero-crossing than other
amplitude-based methods such as the power spectrum and DFA (see also
Fig.~1).

Similar to the literature$^8$, state dependence of the EEG fractal
is observed: the $\nu$ exponent is larger in EO than in EC. In
addition, $\nu$ and $R_\alpha$ follows an inverse relation in both
EC and EO (Fig.~5a). If one tentatively compares to the Hurst model
using $h=2-\nu$, this result means a positive correlation between
the EEG fractal background scaling and the underlying $\alpha$
rhythm.

\subsection{Alpha Interval and Organization of Complex Brain Dynamics}

The inverse relationship shown in Fig.~5a further suggests a relationship
between the scaling property of the EEG fractal background and the
$\alpha$ dynamics.

Consider first the zero-crossing points of the EEG fractal background
in the real line: $\IF=\{t_1,t_2,\cdots\}$ where $t_i$ locates the
end points of the CTI defined in $\F$. The set $\IF$ can be obtained
by intersecting the fractal portion of the EEG graph ${\cal X}_\F$
with the zero axis. Let ${\rm dim}_\IF$ be the box-counting dimension
for $\IF$. It is known from geometry that ${\rm dim}_{{\cal X}_\F}+1-
2={\rm dim}_\IF$ where the `1' on the left-hand-side of this equation
is the dimension of the zero axis. By the H$\ddot{\rm o}$lder
condition of the fractal function$^{19}$, one can show ${\rm dim}_{{
\cal X}_\F}\le 2-(2-\nu)=\nu$. Thus, ${\rm dim}_\IF\le\nu-1$.

Similarly, one can define the zero-crossing points of the $\alpha$
dynamics. In principle, this is obtained by intersecting the portion
of the $\alpha$ oscillation in the EEG graph with the zero axis. In
practice, we use the $A_i$'s in (1a) to approximate the $\alpha$
interval $L_i=\sum_{k=l_i}^{m_i}\tau_k$ where the $L_i$ measures the
``size" of the $i^{\rm th}$ $\alpha$ event as it emerges in the
foreground of the EEG. Unlike the $\tau_{l_i}$ in (1a), which is
narrowly distributed in the $\alpha$ rhythm range$^{18}$, $L_i$ can
in general cover a much wider range of values. The end points
locating the $\alpha$ intervals define the zero-crossing points $\IA
=\{t_1,t_2,\cdots\}$.

Given the EEG fractal background established in the last section, we
conjecture the same for the $\alpha$ interval and a power law PDF
$p(L)\sim L^{-\phi}$. Denote the box-counting dimension of $\IA$ by
${\rm dim}_\IA$; i.e., $n(T)\sim T^{\ {\rm dim}_\IA}$ where $n(T)$
is the number of zero-crossing points over the period $T$. The power
law exponent $\phi$ can be connected to ${\rm dim }_\IA$ via the
relation$^{14}$: $n(T)=T-n(T)\int^Tp(L)LdL$ where $n(T)\int^Tp(L)LdL$
on the right-hand-side estimates the average length of all $\alpha$
intervals over the period $T$. Substituting the power law $p(L)\sim
L^{-\phi}$ yields ${\rm dim}_\IA=\phi-1$.

Since the point sets $\IF$ and $\IA$ are embedded in the (1D) real
line (time axis) and $\IF\cap\IA=\emptyset$, one has 
$$0={\rm dim}_\IF+{\rm dim}_\IA-1\le(\nu-1)
+(\phi-1)-1.$$
Hence, we have the inequality that connects the power law exponents
of the EEG fractal background and the length of the $\alpha$
rhythm:
$$\nu+\phi\ge 3.
\eqno(3)$$

Due to the little $\alpha$ rhythm in our EO data sets, the estimation
of $p(L)$ was suffered from poor statistics. For the EC data sets,
both power law $p(L)$ and (3) are verified (Fig.~5b). We will thus
focus only on the EC data in this section. In Fig.~5c, an interesting
transition is shown from the inequality (3) in $\alpha$ moderate
cases (S1, S5, S6) to almost equality $\nu+\phi\sim 3$ in $\alpha$
dominant cases (S2, S3, S4). For $\nu+\phi\sim 3$ ($\alpha$ dominant)
the numerical values $\nu\sim 1.25$ and $\phi\sim 1.75$ are estimated.
For $\nu+\phi>3$ ($\alpha$ moderate), larger $\nu$ and $\phi$ are
estimated.

The case of $\phi\sim 1.75$ in the $\alpha$ dominant EEG worths
further discussion. In their model study of multi-trait evolution$
^{14}$, Boettcher and Paczuski (BP) predicted that evolution is a
self-organized critical (SOC) process consisting of quiescent
periods of all sizes, interspersed by short intervals of mutation
event. These authors derived a similar power law scaling exponent
of 1.75 for the distribution of the quiescent period and pointed
out that their multi-trait model belongs to a new universality
class. This universality condition prompted us to make the
comparison of these two different phenomenologies. With the
estimated $\phi\sim 1.75$, it is tempting to compare the $\alpha$
interval with the quiescent period in the BP model. This leads to
the hypothesis of SOC dynamics in the ``resting" state of the
cortex.

\section{Discussion and Conclusion}
\bigskip

The EEG background fluctuation coexisting with varying degrees of
$\alpha$ rhythm is studied. It is important to note that our results
directly address these two prominent features of the brain dynamics,
rather than the EEG fluctuation in the $\alpha$ frequency band$^{7,8}$.
Compared to other amplitude-based methods, we show that zero-crossing
is more effective for studying the scaling in $\alpha$ dominant EEG.

Our main result is the evidence and characterization of the
coupling between the EEG background fluctuation and the $\alpha$
rhythm. Our findings can be summarized in two points:

\noindent (a) An inverse relationship between the EEG background
scaling and the strength of $\alpha$ rhythm is observed, with a
larger $\nu$ exponent in EO for all subjects. Using the Hurst model
tentatively ($h=2-\nu$), this implies a larger scaling exponent in
EC compared to EO, or the trend towards more anti-persistent
fluctuation in the $\alpha$ dominant brain state in EO.

Similar state dependence propery has been reported in the past$^{3,
8}$. Stam and de Bruin$^8$ found similar result based on the
correlation between the $\alpha$ band desynchronization and the
decrease of the (DFA) scaling exponent in EO. An inverse relation
was also reported by Moosmanns et al.$^3$ based on the blood
oxygenation level dependence contrast as a measure of the brain
metabolic activity. These authors concluded a dominant $\alpha$
rhythm associated with metabolic deactivation and desynchronization,
a view may further be supported by the increase of local cortical
activity$^{20}$ and information processing during EO$^{21}$.

\noindent (b) The inequality (3) characterizes the organization of
the coupling. Since it is arrived based solely on the set-theoretic
arguments, it is plausible that similar equations may exist between
the EEG fractal background and other brain rhythms such as the
$\theta$, $\delta$, $\beta$, $\gamma$ waves.

The observed transition from $\nu+\phi>3$ to $\nu+\phi\sim 3$ in
$\alpha$ dominant EEG, and the coincidence with the BP dynamics,
imply a SOC state in the $\alpha$ dominant EEG. Hence, a strong
$\alpha$ rhythm and the corresponding background fluctuation may
represent two perspectives of the same dynamics. In general, this
observation is in agreement with the suggestion of SOC of the
EEG fractal dynamics$^{4,7,8}$. However, our result differs in
that we find indication of SOC only at the $\alpha$ dominant
brain state, based mainly on the universality of the BP dynamics.
For subjects showing little to moderate $\alpha$ rhythm, the
unversality condition is no longer matched. We are not able to
determine if a different universality class may be involved in
these cases, nor can we ascertain a different theory for the
observed fractal dynamics. Further studies on a larger population
size and different physiological states are necessary to provide
answers for these questions. They are the future work currently
underway.

\bigskip
\noindent Acknowledgment
\bigskip

The authors would like to acknowledgment supports from Natural Science
and Engineering Research Council of Canada.

\bigskip
\noindent Reference
\bigskip

\noindent [1] H. Berger, Arch Psychiat Nervenkr, {\bf 87}, 52
(1929); ED. Adrain and BH. Matthews, Brain, {\bf 57} 355 (1934).

\noindent [2] T. Inouye, K. Shinosaki, A. Yagasaki, A. Shimizu, Clin.
Neurophysiol, {\bf 63}, 353 (1986); PL. Nunez, BM. Wingerier, RB
Silberstein, Human Brain Mapping, {\bf 13}, 125 (2001).

\noindent [3] M. Moosmann, P. Ritter, I. Krastel, A. Brink, S. Thees,
F. Blankenburg, B. Taskin, H. Obrig, A. Villringer, Neuroimage, {\bf
20}, 145 (2003); RI Goldman, JM. Stern, J. Engel Jr., MS. Cohen,
Neuroreport, {\bf 13}, 2487 (2002).

\noindent [4] WJ. Freeman, Clin Neurophysiol 115, 2089-2107 (2004).

\noindent [5] PA. Watters, Int J Syst Sci, {\bf 31}, 819 (2000).

\noindent [6] R. Hwa and T. Ferree, Phys Rev E, {\bf 66}, 021901 (2002);

\noindent [7] CJ. Stam, EA. de Bruin, Human Brain Mapping, {\bf 22},
97 (2004).

\noindent [8] K. Linkenkaer-Hansen, V.V. Nikulin, J.M. Palva, K.
Kaila, R. J. Ilmoniemi, J Neurosci, {\bf 15}, 1370 (2001).

\noindent [9] T. Murata, T. Takahashi, T. Hamada, M. Omori, H. Kosaka,
H. Yoshida, Y. Wada, Neuropsychobiol, {\bf 50}, 189 (2004).

\noindent [10] see; e.g., H. Otzenberger, C. Simon, C. Gronfier, G.
Brandenberger, Neurosci Lett, {\bf 229}, 173 (1997); J. Ehrhart,
M. Toussaint, C. Simon, C. Gronfier, R. Luthringer, G. Brandenberger,
Sleep, {\bf 111}, 940 (2000).

\noindent [11] C.-K. Peng, S. Havlin, H.E. Stanley, A.L. Goldberger,
Chaos, {\bf 5}, 82 (1995).

\noindent [12] J. Gaillard, Neuropsychobiol. 14, 210 (1987); B.
Elizabeth et al., Sleep Res Online, {\bf 3}, 113 (2000); RM.
Rangayyan, Biomedical Signal analysis, Wiley-IEEE Press (2001).

\noindent [13] PA. Watters and F. Martin, Biol Psych, {\bf 66}, 79
(2004).

\noindent [14] S. Maslov, M. Paczuski, P. Bak, Phys Rev Lett {\bf
17}, 2162 (1994); M. Paczuski, P. Bak, S. Maslov, Phys Rev Lett
{\bf 74}, 4253 (1995); S. Boettcher and M. Paczuski, Phys Rev Lett
{\bf 76}, 348 (1996).

\noindent [15] H.J. Jensen, Self-Organized Criticality, Cambridge
Univ. Press, Cambridge (1998).

\noindent [16] SO. Rice, Bell Syst Tech J, {\bf 24}, 46 (1945); B.
Derrida, V. Hakim, R. Zeitak, Phys Rev Lett, {\bf 77}, 2871 (1996).

\noindent [17] M. Ding, W. Yang, Phys Rev E, {\bf 52}, 207 (1995).

\noindent [18] The $\alpha$ wave crosses the zero axis twice per
cycle. Assuming the mean frequency $\sim$10 Hz, the CTI is
$\sim$1/10/2 second.

\noindent [19] See Prop. 4.2 in K. Falconer, {\it Fractal Geometry},
John Wiley \& Sons, New York (1995). For fBm, it is possible to
achieve equality using potential theoretic approach and the Gaussian
distribution of the increment.

\noindent [20] G. Pfurtscheller and A. Aranibar, Electroenceph Clin
Neurophysiol, {\bf 42}, 817 (1997).

\noindent [21] CJ. Stam, M Breakspear, AM. Van Cappellen van Walsum,
BW. van Dijk, Human Brain Mapping, {\bf 19}, 63 (2003).

\vfill\eject
\bigskip
\noindent Figure Caption

\bigskip
\noindent Fig. 1 (a) EEG records with moderate (top) and strong
(bottom) $\alpha$ rhythm, and the corresponding (b) power
spectral density functions and (c) the DFA result$^{11}$ $F(l)
\sim l^b$. The results for moderate (strong) $\alpha$ rhythm
are the top (bottom) curves in (b) and (c). They correspond
to, respectively, subjects S4 and S5 in eyes closed (Fig.~4).
The solid line in (b) marks the frequency 10 Hz ($\log(10)\sim
2.3$). The solid lines in (c) have the slopes $\sim$1.21 and
$\sim$0.51, respectively.

\bigskip 
\noindent Fig. 2 (a) An example of the CTI of $A_h(t)$, $h=0.3$.
(b) Log-log plot of $p(\tau)$ for $A_{0.3}(t)$ (top) and $A_{0.8}
(t)$ (bottom) before (open circles) and after (crosses) deleting
${\cal A}$ (see (1)). The axes are arbitrary. The solid lines are
drawn with the theoretical slope $-1.2 (=0.8-2)$ and $-1.7 (=0.3-
2)$. The filled circles describe the zero-crossing PDF of a
gaussian white noise, where no power law can be claimed$^{13}$.

\bigskip
\noindent Fig. 3 (a) A segment of the synthetic EEG. Two fractal
periods are highlighted by horizontal bars in $161.5\sim 162.5$.
(b) A segment of the set ${\cal C}$. Note the concentration of
$\tau_i\sim 0.05$ sec. (c) The set ${\cal C}\backslash\A$ where
$\A$ is defined by (1). The horizontal lines mark the levels of
$\tau_u\sim\exp(-2.5)$ and $\tau_l\sim\exp(-5)$. (d) Log-log plot
of $p(\tau)$ before (connected dots) and after (heavy solid line)
deleting ${\cal A}$. The light solid line has the theoretic
slope, -1.9, from the fractal component in synthetic data ($A_{
0.1}(t)$). The vertical line marks the 10 Hz oscillation$^{18}$
($\tau=1/20,\log(\tau)\sim 2.9$).

\bigskip
\noindent Fig. 4 (a) The $R_\alpha$ measure for the six subjects.
(b) Log-log plot of $p(\tau)$ for subject S4 in EC showing
dominant $\alpha$ rhythm. A segment of the EEG for this subject
has been shown in Fig.~1a. (c) Log-log plot of $p(\tau)$ for a
subject in EO showing little $\alpha$ rhythm (S6 in (a)). The
close (open) circles show the $p(\tau)$ before (after) deleting
${\cal A}$ and the solid lines are the regression lines ($\tau_l
\sim\exp(-5)$ and $\tau_u\sim\exp(-1)$).

\bigskip
\noindent Fig. 5 (a) The inverse relationship between $\nu$ and
$R_\alpha$ in EO (open circles) and EC (solid circles). The
subject index (used in Fig.~4) is given next to the symbol. (b)
Log-log plots of $p(\tau)$ and $p(L)$ for subjects showing
dominant $\alpha$ rhythm in EC: S2 (in asterik and circle, resp.),
S3 (in plus and triangle, resp.), S4 (in cross and square, resp.)
Solid lines with slopes -1.25 and -1.75 are drawn. (c) The plot
of $\nu$ (solid circle), $\phi$ (open square) and $\nu+\phi$
(cross) for subjects in EC. $R_\alpha$ in Fig.~4a is reproduced
and referenced to the left $y$-axis.

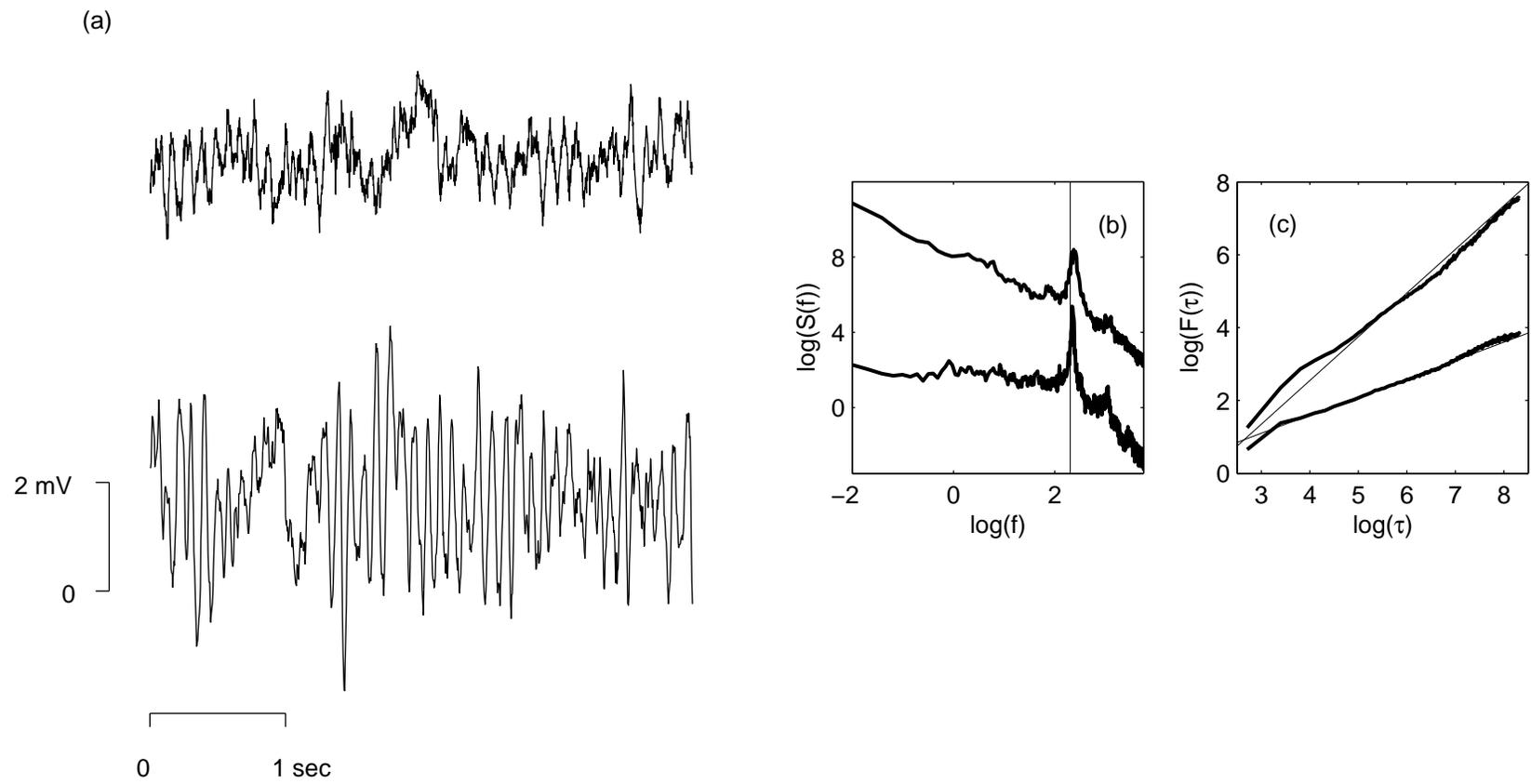
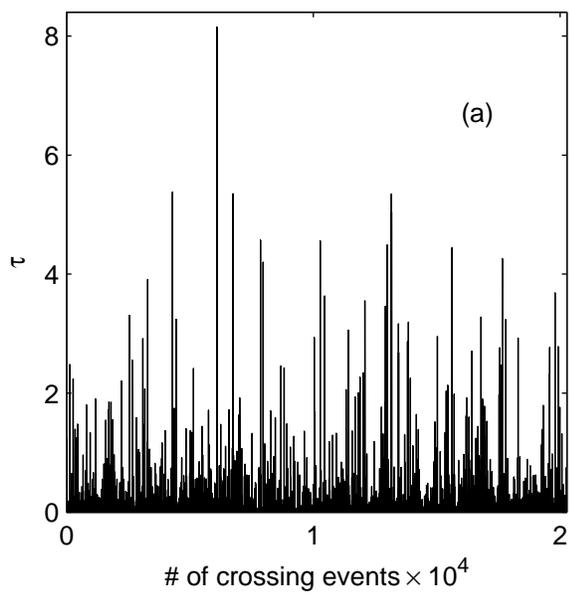
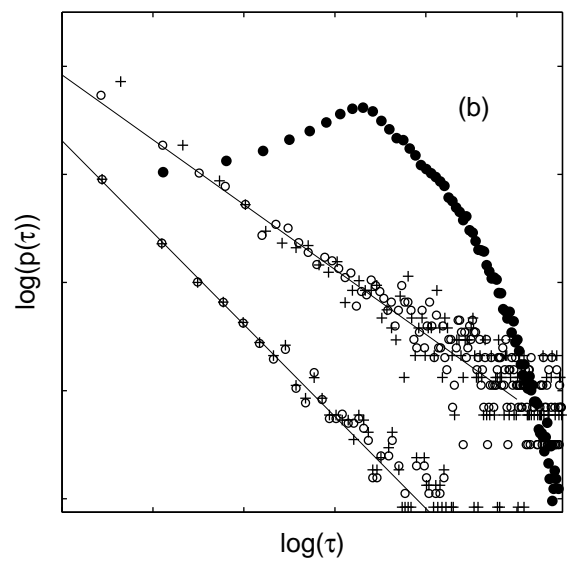
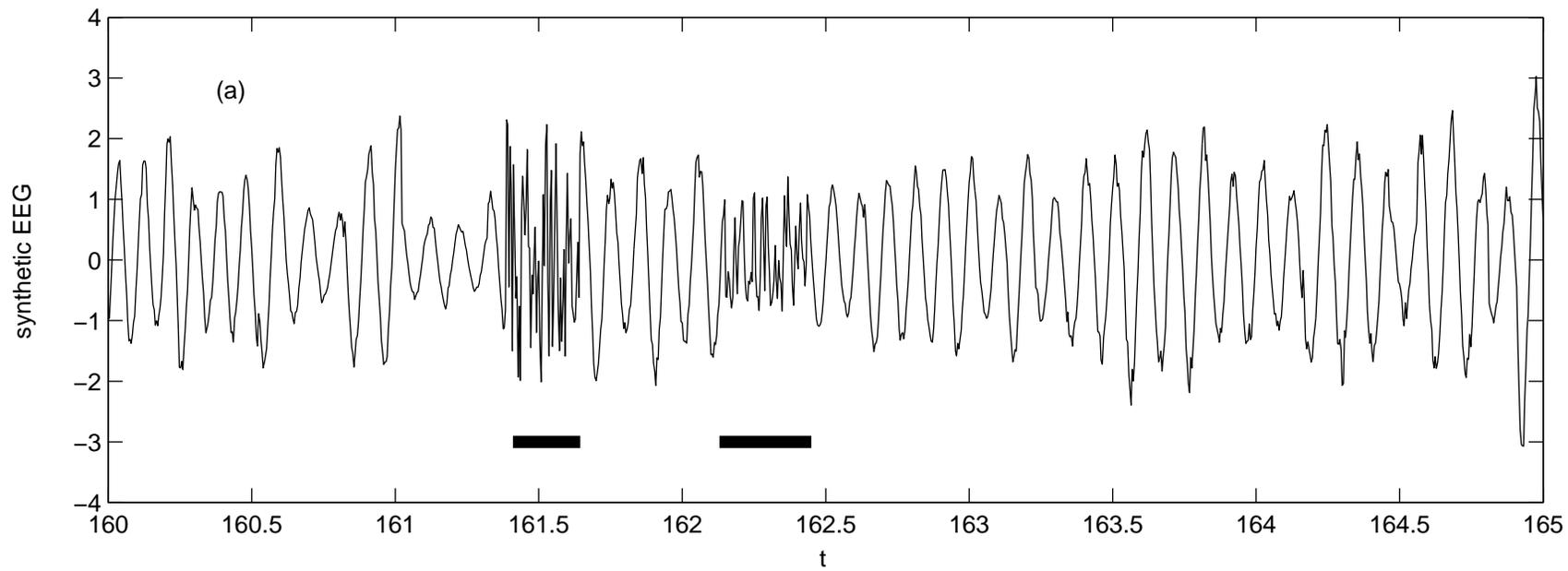
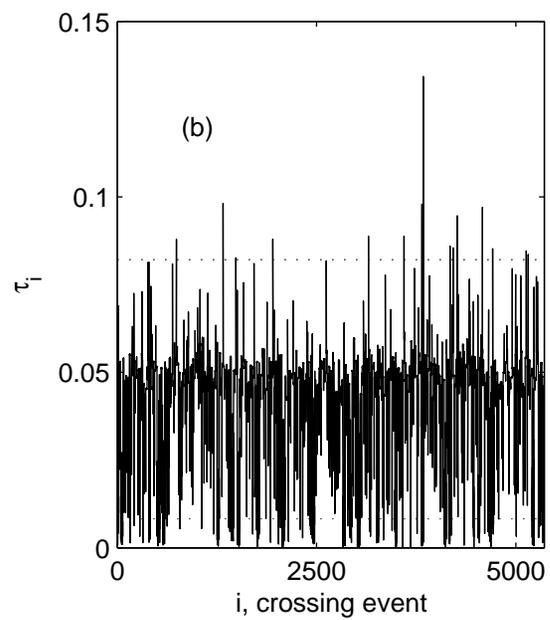
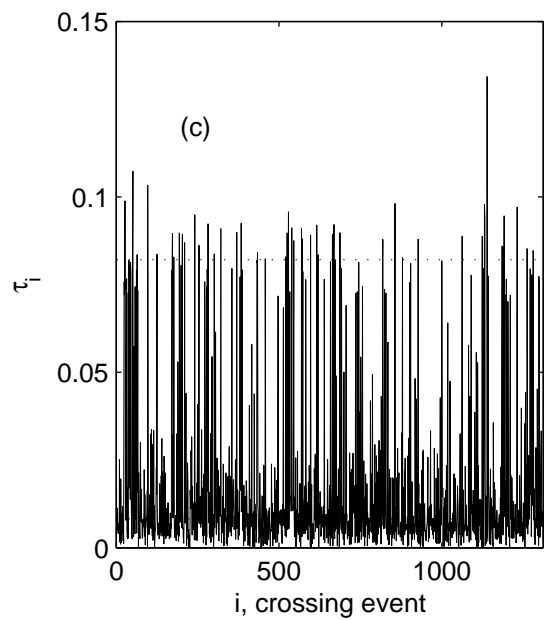
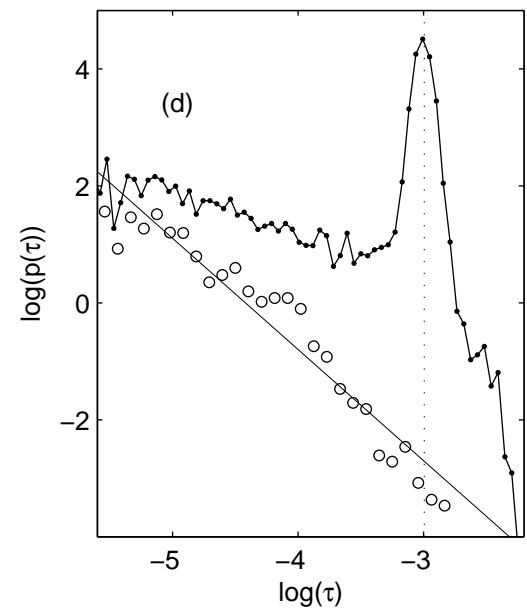
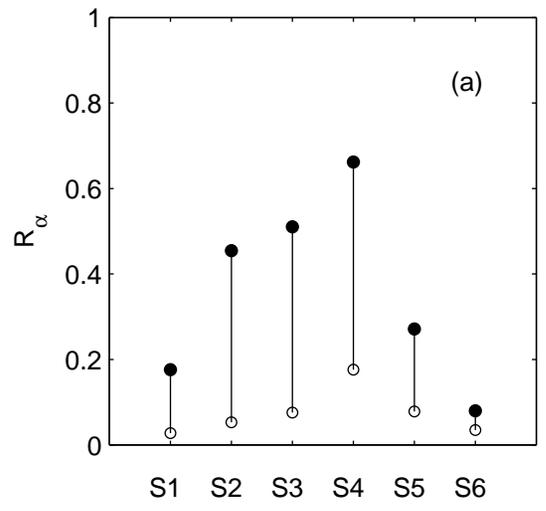
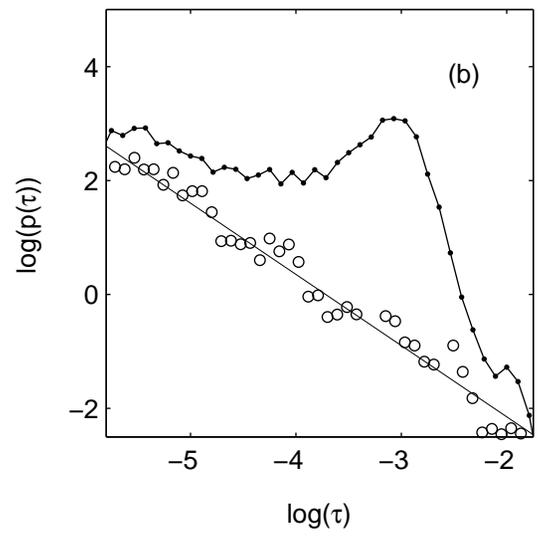
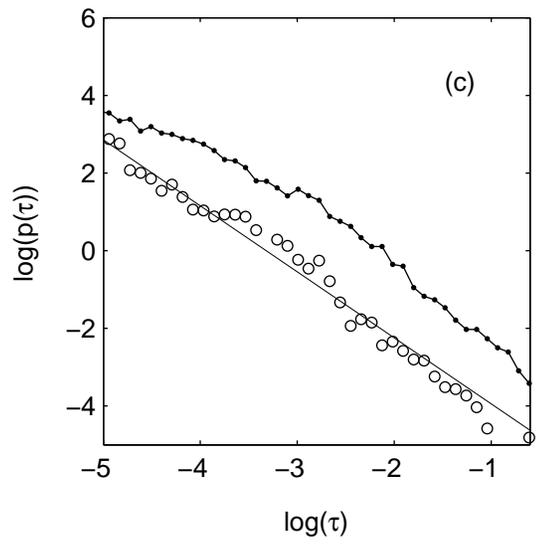
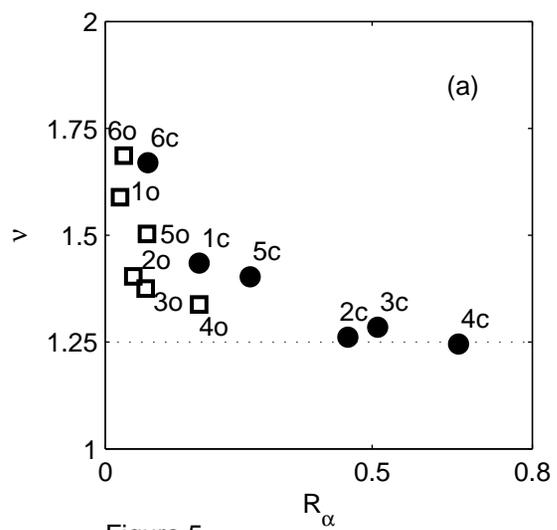
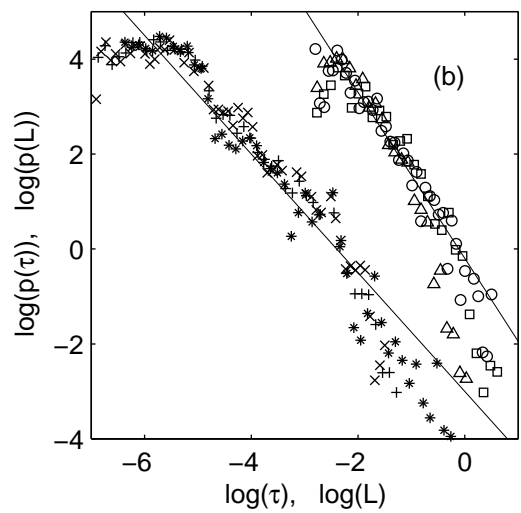
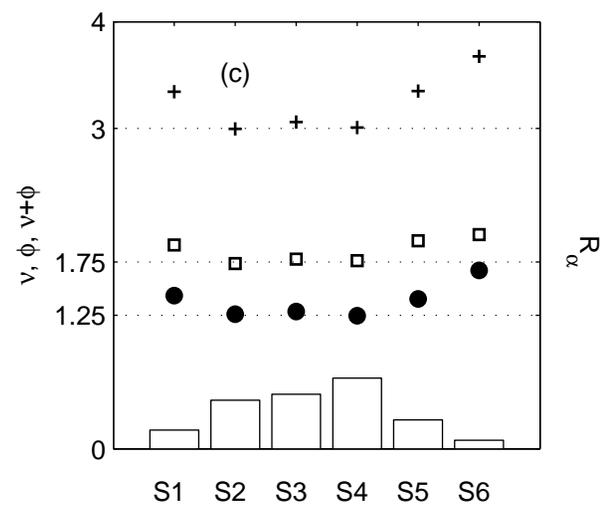
\end{document}